\documentclass[12pt, reqno]{amsart}

\usepackage{amssymb} 

\usepackage{graphicx}

\newcommand{\wt}{\widetilde}

\newtheorem{dfn}{Definition}
\newtheorem{prop}{Proposition}
\newtheorem{thm}{Theorem}

\newcommand{\wh}[1]{\widehat{#1}}
\newcommand{\ov}[1]{\overline{#1}}

\begin{document}

\title[Kochen-Specker on three qubits]{New examples of Kochen-Specker type configurations on three qubits}

\author[A. E. Ruuge]{Artur E. Ruuge}
\address{
Department of Mathematics and Computer Science, 
University of Antwerp, 
Middelheim Campus Building G, 
Middelheimlaan 1, B-2020, 
Antwerp, Belgium
}
\email{artur.ruuge@ua.ac.be}

%PACS numbers: 03.65.Aa, 03.65.Ud

\begin{abstract}
A new example of a saturated Kochen-Specker (KS) type configuration of 64 rays 
in 8-dimensional space (the Hilbert space of a triple of qubits) is constructed. 
It is proven that this configuration has a tropical dimension 6 and that 
it contains a critical subconfiguration of 36 rays. 
A natural multicoloured generalisation of the Kochen-Specker theory is given 
based on a concept of an entropy of a saturated configuration of rays.  
\end{abstract}

\maketitle

\section{Introduction}

The purpose of the present paper is 
to introduce some new theoretical concepts 
related to the Bell-Kochen-Specker theory, 
and to illustrate them on a new example of a 
\emph{saturated} Kochen-Specker (KS) configuration 
in eight-dimensional space (the Euclidean space of three qubits). 
The saturated configurations are of special interest due to their symmetry properties.  
The present configuration consists of $64 = 2^{6}$ rays and it is 
the smallest 
non-trivial %%%%%%%%%%
\emph{saturated} KS configuration known at this moment in dimension $8$. 

The KS configurations are sometimes termed ``non-colourable'' \cite{ZimbaPenrose} 
and the terminology associated to them can slightly vary from author to author. 
The original construction of S.~Kochen and E.~P.~Specker can be found in \cite{KochenSpecker} 
(a configuration of $117$ rays in three dimensions), 
and a motivation behind their work can be traced back to the results of J.~S.~Bell \cite{Bell} 
(the Bell inequalities in quantum mechanics).

The KS configurations are of interest for the foundations of quantum mechanics 
\cite{AbramskyHardy, BancalGisinPironio, IshamLinden, Smith, Vourdas, Wilce} since 
they do not refer directly to the concept of \emph{probability}. 
An additional reason to study such configurations is due to the links 
to some issues of quantum gravity 
\cite{ButterfieldIsham1, ButterfieldIsham2, HamiltonIshamButterfield, Isham, IshamButterfield, IshamLinden}. 
On the other hand, they are of interest in the quantum information science (QIS). 
The literature about QIS is quite extensive in our days. 

The example of the saturated KS configuration constructed in the present paper (64 rays) 
is similar to the configurations described, for instance, in 
\cite{CabelloEstebaranzGarcia-Alcaine, Mermin, Planat} 
in the sense 
that the coordinates of the the vectors representing the rays can be chosen to be $0$, $1$, or $-1$. 
In a recent paper \cite{WaegellAravind} M.~Waegell and P.~K.~Aravind 
point out that there exists a rather large class of KS proofs related to the Pauli group on $N$ qubits. 
The configurations of rays underlying such proofs can be described by vectors 
with coordinates $0$, $\pm 1$, and $\pm i$, where $i$ is the imaginary unit. 
I observe that the saturated configuration of 64 rays mentioned admits such a proof. 
I also compute a \emph{critical} subconfiguration $\mathcal{N} \subset \mathcal{M}$ in it 
(i.e. the KS configuration $\mathcal{N}$ which cannot be made smaller by deleting the rays)
which happens to contain $36$ rays 
(the same number as in the critical configuration of \cite{KernaghanPeres}, 
but the configurations are \emph{not} isomorphic).

In the present paper I consider \emph{three} types of KS configurations: 
saturated, critical, and \emph{tropical}. 
The first two types (saturated and critical) have already appeared in the literature 
\cite{Larsson, PavicicMerletMcKayMegill, Ruuge2, RuugeFVO1, ZimbaPenrose},
and the term ``tropical'' is new. 
It is a ``critical'' configuration, but in a different sense, which 
is explained in the next section.  
It is being used as an intermediate step to find a ``small'' critical subconfiguration of $\mathcal{M}$. 
There is a sequence of inclusions: 
\begin{equation*} 
\mathcal{N} \subset \mathcal{T} \subset \mathcal{M}, 
\end{equation*}
where $\mathcal{M}$ is a saturated KS configuration (64 rays), $\mathcal{T}$ is tropical (48 rays), and 
$\mathcal{N}$ is critical (36 rays).  
The configuration $\mathcal{N}$ is obtained from $\mathcal{T}$ by deleting the rays. 

Small critical KS configurations are more easy to test in experiment 
(they must be less expensive). 
For instance, there exists an example of a KS configuration 
in four dimensions consisting of just 18 rays 
\cite{CabelloEstebaranzGarcia-Alcaine} 
which has been tested experimentally \cite{Experiment1, Experiment2, Experiment3, Experiment4}. 
The dimension four can be perceived as the dimension of the Euclidean space corresponding to a pair of qubits ($4 = 2 \times 2$). 
In the present paper the dimension is eight, which is the dimension of a three-qubit system ($8 = 2 \times 2 \times 2$). 
A triple of qubits is a rather special quantum object on its own. 
To observe the EPR-type correlations \cite{EPR} one may work with only two qubits, but to observe the \emph{really} weird
features of quantum theory it is better to have at least three qubits. 
A recent experimental realisation of a triple of qubits on a ``single chip'' has been reported in \cite{IBM}.  
In another work, 
it has recently been shown \cite{VertesiBrunner} that 
a triple of qubits suffices to refute the conjecture of A.~Peres about quantum nonlocality and entanglement distillability. 
One can observe Berry's phase \cite{Berry} on a triple of qubits and describe it in terms of quantum groups \cite{HuWuXue}. 
There exists a non-trivial link between the $E_8$-root system and the theory of three qubits found in \cite{Ruuge2, RuugeFVO1}. 
It turns out that the rays represented by the roots of an $E_8$-type Lie algebra yield an example of a \emph{saturated} Kochen-Specker type configuration and 
that this fact can be generalised in several different ways. One can obtain an infinite family of orthoalgebras \cite{RuugeFVO2} 
starting from the $E_8$-root system 
which can be of interest in some approaches to quantum gravity. 
% advocated in \cite{Isham}. 
On the other hand one can scale up and deform this example on multiple qubits \cite{Ruuge1}, 
as well as one can consider other root systems \cite{Ruuge2}.

The present paper focuses on the case of a three qubit system, 
although other important KS examples are known in other dimensions 
\cite{AravindLee-Elkin, BadziagBengtssonCabelloPitowsky, Cabello, CabelloEstebaranzGarcia-Alcaine, 
HarveyChryssanthacopoulos, KochenSpecker,  
Peres, Planat, PlanatSaniga, Ruuge1, Ruuge2, 
WaegellAravind, ZimbaPenrose}. 
In the beginning I introduce several new definitions and then I test them on the 
saturated configuration $\mathcal{M}$ (64 rays) mentioned above. 
The proof that the configuration is of a Kochen-Specker type can be given analytically (see Appendix A), 
or checked on a computer. 
The supplementary files mentioned in the main text can be found on 
arXiv or on my personal webpage. 
It is worth to draw attention to the concept 
of an \emph{entropy} of a configuration of rays: 
a Kochen-Specker colouring corresonds to a zero entropy $S = 0$. 
It is possible to consider more general colourings with $S > 0$. 
This naturally leads to a \emph{multicoloured} generalisation of 
the Kochen-Specker theory, which is not immediately visible in other approaches (KS proofs in terms of operators). 
I mention a couple of examples related to multi-colourings, but I leave the rest for another paper.

\section{Terminology and notation}

Throughout the paper, 
\begin{equation*} 
[n] := \lbrace 0, 1, \dots, n - 1 \rbrace, 
\end{equation*}
where $n = 1, 2, \dots$. 
It is a little more convenient to start counting from $0$ rather than from $1$ 
having in mind the C programming language.

Let $\mathcal{M}$ be a \emph{finite} collection of rays in a Euclidean space $\mathbb{C}^d$. 
A Kochen-Specker (KS) colouring on $\mathcal{M}$ is a function $f: \mathcal{M} \to \lbrace 0, 1 \rbrace$, such that 
1) for any $x, y \in \mathcal{M}$, if $x$ and $y$ are orthogonal, then at least one of the values $f (x)$ or $f (y)$ is zero; 
2) if $x_0, x_1, \dots, x_{d - 1} \in \mathcal{M}$ is a set of mutually orthogonal rays, then $\sum_{i = 0}^{d - 1} f (x_i) = 1$. 
The collection $\mathcal{M}$ is termed a \emph{KS configuration} of rays iff it does not admit a KS colouring. 
A KS configuration $\mathcal{M}$ is termed \emph{critical} iff it can not be made smaller by deleting the rays, 
i.e. iff for any $x \in \mathcal{M}$ the collection $\mathcal{M} \backslash \lbrace x \rbrace$ admits a KS colouring. 
A collection of rays $\mathcal{M}$ (not necessary a KS configuration) is termed \emph{saturated} iff 
every collection $x_0, x_1, \dots, x_{k - 1} \in \mathcal{M}$ of $k$ mutually orthogonal rays, where $k = 1, 2, \dots, d - 1$, 
can be extended by $x_{k}, x_{k + 1}, \dots, x_{d - 1} \in \mathcal{M}$ to a collection of $d$ mutually orthogonal rays 
($d$ is the dimension of space). 

It is natural to associate to $\mathcal{M}$ an undirected graph as follows: 
the vertices are the elements of $\mathcal{M}$; 
a pair of vertices $x, y \in \mathcal{M}$ is connected with an edge iff $x \perp y$. 
Let us term this graph the \emph{orthogonality graph} of $\mathcal{M}$ 
and denote it $\Gamma (\mathcal{M})$. 
A \emph{clique} in $\Gamma (\mathcal{M})$ of size $k$ is a 
collection of $k$ distinct vertices $x_0, x_1, \dots, x_{k - 1} \in \mathcal{M}$  
such that every vertex $x_i$ is connected to every other vertex $x_j$ with an edge, $i \not = j$, $i, j \in [k]$. 
%%%%%%%%%%
A \emph{maximal clique} in $\Gamma􏰞(M)$ is a 
clique of the largest possible size $\omega \leqslant d$ ($d$ is the dimension of the Euclidean space).
%%%%%%%%%%
Denote the set of all cliques of size $k$ as $\mathcal{P}_{\perp}^{(k)} (\mathcal{M})$, $k \in \mathbb{Z}_{> 0}$. 
A \emph{clique covering} of the graph $\Gamma (\mathcal{M})$ is a collection of cliques 
$U_i \in \mathcal{P}_{\perp}^{(m_i + 1)} (\mathcal{M})$, $m_{i} \in [d]$, $i \in [n]$, 
such that $\cup_{i \in [n]} U_i = \mathcal{M}$. 
A \emph{clique partition} of the graph $\Gamma (\mathcal{M})$ is a clique covering $\lbrace U_i \rbrace_{i \in [n]}$ 
such that $U_i \cap U_j = \emptyset$, $i \not = j$, $i, j \in [n]$. 
An \emph{anticlique} in $\Gamma (\mathcal{M})$ of size $k$ is a 
collection of $k$ distinct vertices $x_0, x_1, \dots, x_{k - 1} \in \mathcal{M}$  
such that every vertex $x_i$ is \emph{not} connected to any other vertex $x_j$ with an edge, $i \not = j$, $i, j \in [k]$. 
A maximal anticlique 
can have a size greater than $d$. 
Denote the set of all anticliques of size $k$ as $\mathcal{A}_{\perp}^{(k)} (\mathcal{M})$, $k = 1, 2, \dots $. 

\begin{dfn} 
Configurations of rays $\mathcal{M}$ and $\mathcal{M}'$ are termed \emph{isomorphic} iff 
their orthogonality graphs $\Gamma (\mathcal{M})$ and $\Gamma (\mathcal{M'})$ are isomorphic. 
\end{dfn}
The concept of an isomorphism of configurations should not be confused with an isomorphism of \emph{proofs} 
that the configurations do not admit a KS colouring (the \emph{KS proofs}). 
A KS proof is given by a \emph{hypergraph} with a set of vertices $V \subset \mathcal{M}$, 
and a set of hyper-edges $U_0, U_1, \dots, U_{n - 1} \in \mathcal{P}_{\perp}^{(d)} (\mathcal{M})$. 
The property that ensures that $\mathcal{M}$ is a KS configuration is as follows: 
one cannot construct a function $f: \mathcal{M} \to \lbrace 0, 1 \rbrace$ such that 
for every $i \in [n]$ there is a unique $x \in U_i$ assigned with $f (x) = 1$. 
Such hyper-graphs (in case $d = 4$) are studied in the approach of 
\cite{McKayMegillPavicic, MegillKresimirWaegellAravindPavicic, PavicicMcKayMegillFresl, PavicicMerletMcKayMegill} 
using some algorithms from \cite{McKay}.
An \emph{isomorphism of KS proofs} is an isomorphism of these hyper-graphs. 

\begin{dfn} 
A \emph{signature} of a configuration of rays $\mathcal{M}$ is a pair of functions $(f, g)$, 
where $f: \mathbb{Z}_{> 0} \to \mathbb{Z}$, $k \mapsto \# \mathcal{P}_{\perp}^{(k)} (\mathcal{M})$, and 
$g: \mathbb{Z}_{> 0} \to \mathbb{Z}$, $k \mapsto \# \mathcal{A}_{\perp}^{(k)} (\mathcal{M})$. 
\end{dfn}
If the signatures of configurations $\mathcal{M}$ and $\mathcal{M}'$ are different, 
then the configurations cannot be isomorphic. 

\begin{dfn} 
Let $\mathcal{M}$ be a set of rays in a $d$-dimensional Euclidean space. 
Let $\lbrace U_i \rbrace_{i \in [n]}$ be a collection of cliques 
$U_i \in \mathcal{P}_{\perp}^{(m_i + 1)} (\mathcal{M})$, $m_i \in [d]$, $i \in [n]$. 
An \emph{anticlique section} of this collection is a function 
$f: \cup_{i \in [n]} U_i \to \lbrace 0, 1 \rbrace$, 
such that for every $i \in [n]$ there exists a unique $x \in U_i$ such that $f (x) = 1$. 
\end{dfn}
Notice that 
if $f$ is an anticlique section of $\lbrace U_i \rbrace_{i \in [n]}$, then 
$f^{- 1} (\lbrace 1 \rbrace) \in \mathcal{A}_{\perp}^{(k)} (\mathcal{M})$, where $k \leqslant n$. 
The cardinality $k$ is not necessary equal to $n$ since 
the cliques $U_i$, $i \in [n]$, can have non-empty intersections. 
It is also important to point out that even if $\lbrace U_i \rbrace_{i \in [n]}$ admits an anticlique section, 
the union $\cup_{i \in [n]} U_i$ can still be a KS configuration, since there can be other orthogonalities 
between the rays which are not described by $U_i$, $i \in [n]$. 

\begin{dfn}
Let $\mathcal{M}$ be a KS configuration of rays in a $d$-dimensional Euclidean space, 
such that every pair of orthogonal rays $x, y \in \mathcal{M}$, $x \perp y$, 
is contained in a maximal clique $U \in \mathcal{P}_{\perp}^{(d)} (\mathcal{M})$.  
The smallest integer $n$ such that there exists a collection of maximal cliques 
$\lbrace U_i \rbrace_{i \in [n]}$, $U_i \in \mathcal{P}_{\perp}^{(d)} (\mathcal{M})$, $i \in [n]$, 
which does not admit an anticlique section, is termed a \emph{tropical dimension} of $\mathcal{M}$. 
Notation: $n = \dim (\mathcal{M})$. 
\end{dfn}
Notice that for the collection $\lbrace U_i \rbrace_{i \in [n]}$ mentioned in the definition, 
where $n = \dim (\mathcal{M})$ is the tropical dimension of $\mathcal{M}$, 
we can conclude that $\mathcal{T} := \cup_{i \in [n]} U_i$ is a KS configuration of rays. 
At the same time, the collection $\lbrace U_i \rbrace_{i \in [n]}$ 
is only a \emph{proper} subset of the hyper-edges of the hyper-graph 
describing a KS proof for $\mathcal{T}$. 
The cliques $U_i$, $i \in [n]$, can even be mutually disjoint. 
In the present paper I compute the tropical dimension of 
the saturated KS configuration $\mathcal{M}$ of 64 rays in 8-dimensional space 
mentioned in the introduction (the answer is $\dim (\mathcal{M}) = 6$).
\begin{dfn} 
Let $\mathcal{T}$ be a KS configuration of rays in a $d$-dimensional Euclidean space, 
such that every pair of orthogonal rays $x, y \in \mathcal{T}$, $x \perp y$, 
is contained in a maximal clique $U \in \mathcal{P}_{\perp}^{(d)} (\mathcal{T})$. 
The configuration $\mathcal{T}$ is termed \emph{tropical} iff 
there exists a covering $\lbrace U_i \rbrace_{i \in [n]}$ of $\mathcal{T}$ by 
$n = \dim (\mathcal{T})$ maximal cliques 
$U_i \in \mathcal{P}_{\perp}^{(d)} (\mathcal{T})$, $i \in [n]$, 
which does not admit an anticlique section. 
\end{dfn}
In the present paper I use a tropical configuration $\mathcal{T} \subset \mathcal{M}$ (48 rays) 
as a starting point to search for a critical subconfiguration of $\mathcal{M}$. 
The output is a critical configuration $\mathcal{N} \subset \mathcal{T}$ consisting of 36 rays.

To describe $\mathcal{M}$ in more detail, it is of interest to 
consider a generalisation of the concept of a KS colouring. 
Let $\mathcal{M}$ be a finite \emph{saturated} collection of rays a Euclidean space of dimension $d$, 
(not necessary a KS configuration). 
A function $f: \mathcal{M} \to [0, 1]$ is termed a \emph{probability weight} on $\mathcal{M}$ iff 
for every $U \in \mathcal{P}_{\perp}^{(d)} (\mathcal{M})$ holds: 
\begin{equation*} 
\sum_{x \in U} f (x) = 1. 
\end{equation*}
Consider an \emph{entropy}: 
\begin{equation*} 
S_{U}^{f} := - \sum_{x \in U} f(x) \log f (x), 
\end{equation*}
for every $U \in \mathcal{P}_{\perp}^{(d)} (\mathcal{M})$. 
If the function $f$ takes only two values $0$ and $1$, then we have a KS colouring of $\mathcal{M}$, 
and all the entropies $S_{U}^{f} = 0$, $U \in \mathcal{P}_{\perp}^{(d)} (\mathcal{M})$. 
It is important to notice that this does \emph{not} 
imply that there is a quantum state with such entropies \cite{Deutsch, WehnerWinter}. 
Associate to every $x \in \mathcal{M}$ an orthogonal projector 
$\wh{\pi}_{x}$ on $x$ (i.e. a quantum observable). 
Every $U \in \mathcal{P}_{\perp}^{(d)} (\mathcal{M})$ corresponds to a maximal set of 
compatible observables $\lbrace \wh{\pi}_{x} \rbrace_{x \in U}$. 
For every quantum state described by a statistical operator $\wh{\rho}$, we can compute  
\begin{equation*} 
S_U [\wh{\rho}] := - \sum_{x \in U} \mathrm{Tr} (\wh{\pi}_{x} \wh{\rho}) \log ( \mathrm{Tr} (\wh{\pi}_{x} \wh{\rho}) ). 
\end{equation*}
In general, even if $\mathcal{M}$ admits a KS colouring $f$, 
one cannot find a quantum state $\wh{\rho}$ such that $S_{U} [\wh{\rho}] = 0$, 
for all $U \in \mathcal{P}_{\perp}^{(d)} (\mathcal{M})$. 

Let us term a probability weight $f: \mathcal{M} \to [0, 1]$ \emph{equientropic} iff 
there exists a constant $S_{0}^{(f)} \in \mathbb{R}$, such that 
$S_{U}^{f} = S_{0}^{(f)}$, for every $U \in \mathcal{P}_{\perp}^{(d)} (\mathcal{M})$. 
Denote the collection of all equientropic probability weights as $\mathcal{E}_{S} (\mathcal{M})$. 
Notice that $\mathcal{E}_{S} (\mathcal{M})$ is not empty, since it contains at least one function: 
$f (x) = 1/ d$, for all $x \in \mathcal{M}$. 
The corresponding entropy $S_{0}^{(f)} = \log (d)$ and this is the maximal possible value of $S_{0}^{(f)}$, 
$f \in \mathcal{E}_{S} (\mathcal{M})$. 
Put 
\begin{equation*} 
S (\mathcal{M}) := \inf \lbrace S_{0}^{(f)} \,|\, f \in \mathcal{E}_{S} (\mathcal{M}) \rbrace. 
\end{equation*}
It is natural to term $S (\mathcal{M})$ an \emph{entropy} of the configuration $\mathcal{M}$. 
The quantity $D (\mathcal{M}) := \exp (S (\mathcal{M}))$ is a certain ``dimension'' describing $\mathcal{M}$ 
which can be termed a  \emph{statistical weight} of $\mathcal{M}$.

The link between the concept of an entropy and the concept of a KS colouring provides 
a non-trivial theoretical insight about the Kochen-Specker topic. 
Take any $f \in \mathcal{E}_{S} (\mathcal{M})$ and denote the range of its values 
$w_0 < w_1 < \dots < w_{m - 1}$, $w_i \in [0, 1]$, $i \in [m]$, $m \leqslant d$. 
For every $U \in \mathcal{P}_{\perp}^{(d)} (\mathcal{M})$, denote $N_{i}^{U}$ the number of 
times $f (x)$ takes the value $w_i$ as $x$ varies over $U$.  
Then we have: 
\begin{equation*} 
\sum_{i \in [m]} N_{i}^{U} = d, \quad 
\sum_{i \in [m]} N_{i}^{U} w_{i} = 1, \quad 
\sum_{i \in [m]} N_{i}^{U} w_{i} \log (w_{i}) = S_{0}^{(f)},  
\end{equation*}
for every  $U \in \mathcal{P}_{\perp}^{(d)} (\mathcal{M})$. 
If all values $w_i$ are \emph{rational}, then 
write  $w_i = q_i/ \Gamma$, where $q_i \in \mathbb{Z}$, 
$i \in [m]$, and assume that $\Gamma \in \mathbb{Z}_{> 0}$ is chosen as small as possible. 
We can perceive $f: \mathcal{M} \to [0, 1] \cap \mathbb{Q}$, 
$f \in \mathcal{E}_{S} (\mathcal{M})$, as a kind of ``multicoloured'' 
generalisation of a KS colouring: $q_i$ can be termed \emph{mixed} colours, and the factors 
present in their factorisations into primes can be termed \emph{primary} colours. 
In a particular case where all $q_i$ are \emph{prime}, $i \in [m]$, 
we obtain just a condition: $N_{i}^{U} = c_i$, where $c_i \in \mathbb{Z}_{> 0}$, $i \in [m]$, for all 
$U \in \mathcal{P}_{\perp}^{(d)} (\mathcal{M})$. 
If we are not interested in a particular value of $S_{0}^{(f)}$, then this motivates the following: 
\begin{dfn} 
Let $\mathcal{M}$ be a collection of rays in a Euclidean space of dimension $d$. 
Let $d = N_0 + N_1 + \dots + N_{s - 1}$ be a partition of $d$ into a sum of $s \geqslant 2$ positive integers, 
$N_0 \geqslant N_1 \geqslant \dots \geqslant N_{s - 1}$. 
A function $h: \mathcal{M} \to \lbrace 0, 1, \dots, s - 1 \rbrace$ is termed a 
\emph{colouring of $\mathcal{M}$ compatible with this partition} iff 
for every clique $U \in \mathcal{P}_{\perp}^{(d)} (\mathcal{M})$ and for every $\alpha \in [s]$ holds: 
$\# \lbrace x \in U \,|\, h (x) = \alpha \rbrace = N_{\alpha}$. 
\end{dfn}
A KS-colouring is just a colouring compatible with the partition $(d - 1, 1) \vdash d$. 
In the present paper I give a pair of examples of colourings compatible with other partitions, and 
an example when such a colouring does not exist.

\section{Saturated Kochen-Specker on three qubits}

The dimension of the Hilbert space of a single qubit is 2. 
The dimension of the Hilbert space $\mathcal{H}$ of a triple of qubits is $d = 8$, $8 = 2 \times 2 \times 2$. 
In this section we describe a new \emph{saturated} Kochen-Specker configuration $\mathcal{M}$ in $\mathcal{H}$ 
consisting of $N = 64$ rays: $\mathbb{C} \psi_0, \mathbb{C} \psi_1, \dots, \mathbb{C} \psi_{N - 1}$. 
Write $\psi_{i}$ as a line of eight numbers (the coordinates of a vector in a selected basis), 
\begin{equation*} 
\psi_{i} = (\psi_{i}^{(0)}, \psi_{i}^{(1)}, \dots, \psi_{i}^{(d - 1)}), 
\end{equation*}
where $i \in [N]$. 
The inner product $\langle -, - \rangle$ on $\mathcal{H} = \mathbb{C}^{d}$ is then given by 
\begin{equation*} 
\langle \psi_i, \psi_j \rangle = 
\sum_{k = 0}^{d - 1} \ov{\psi}_{i}^{(k)} \psi_{j}^{(k)}, 
\end{equation*}
where the bar denotes complex conjugation, $i, j \in [N]$. 
The collection of vectors $\lbrace \psi_{i} \rbrace_{i \in [64]}$ described below has an additional property: 
\begin{equation*}
\psi_{i}^{(k)} \in \lbrace -1, 0, 1 \rbrace, 
\end{equation*}
for $i \in [N]$, and $k \in [d]$, 
so it makes sense to use a special notation. 
We write $\bar 1$ instead of $-1$ and skip the spaces and commas between the coordinates of a vector: 
for example, $\psi = (1 {\bar 1} {\bar 1} 1 0 0 0 0)$ is a vector 
$\psi = (1, -1, -1, 1, 0, 0, 0, 0)$. 
Put: 
\begin{equation}
\begin{array}{lll}
\psi_{0} := (1 {\bar 1} {\bar 1} {\bar 1} {\bar 1} {\bar 1} {\bar 1} 1 ), 
&\psi_{1} := (1 {\bar 1} {\bar 1} {\bar 1} {\bar 1} {\bar 1} 1 {\bar 1} ), 
&\psi_{2} := (1 {\bar 1} {\bar 1} {\bar 1} 1 1 {\bar 1} 1 ), \\
\psi_{3} := (1 {\bar 1} {\bar 1} {\bar 1} 1 1 1 {\bar 1} ), 
&\psi_{4} := (1 {\bar 1} {\bar 1} 1 {\bar 1} {\bar 1} {\bar 1} {\bar 1} ), 
&\psi_{5} := (1 {\bar 1} {\bar 1} 1 {\bar 1} {\bar 1} 1 1 ), \\
\psi_{6} := (1 {\bar 1} {\bar 1} 1 1 1 {\bar 1} {\bar 1} ), 
&\psi_{7} := (1 {\bar 1} {\bar 1} 1 1 1 1 1 ), 
&\psi_{8} := (1 {\bar 1} 1 {\bar 1} {\bar 1} {\bar 1} {\bar 1} {\bar 1} ), \\
\psi_{9} := (1 {\bar 1} 1 {\bar 1} {\bar 1} {\bar 1} 1 1 ), 
&\psi_{10} := (1 {\bar 1} 1 {\bar 1} 1 1 {\bar 1} {\bar 1} ), 
&\psi_{11} := (1 {\bar 1} 1 {\bar 1} 1 1 1 1 ), \\
\psi_{12} := (1 {\bar 1} 1 1 {\bar 1} {\bar 1} {\bar 1} 1 ), 
&\psi_{13} := (1 {\bar 1} 1 1 {\bar 1} {\bar 1} 1 {\bar 1} ), 
&\psi_{14} := (1 {\bar 1} 1 1 1 1 {\bar 1} 1 ), \\
\psi_{15} := (1 {\bar 1} 1 1 1 1 1 {\bar 1} ), 
&\psi_{16} := (1 1 {\bar 1} {\bar 1} {\bar 1} 1 {\bar 1} 1 ), 
&\psi_{17} := (1 1 {\bar 1} {\bar 1} {\bar 1} 1 1 {\bar 1} ), \\
\psi_{18} := (1 1 {\bar 1} {\bar 1} 1 {\bar 1} {\bar 1} 1 ), 
&\psi_{19} := (1 1 {\bar 1} {\bar 1} 1 {\bar 1} 1 {\bar 1} ), 
&\psi_{20} := (1 1 {\bar 1} 1 {\bar 1} 1 {\bar 1} {\bar 1} ), \\
\psi_{21} := (1 1 {\bar 1} 1 {\bar 1} 1 1 1 ), 
&\psi_{22} := (1 1 {\bar 1} 1 1 {\bar 1} {\bar 1} {\bar 1} ), 
&\psi_{23} := (1 1 {\bar 1} 1 1 {\bar 1} 1 1 ), \\
\psi_{24} := (1 1 1 {\bar 1} {\bar 1} 1 {\bar 1} {\bar 1} ), 
&\psi_{25} := (1 1 1 {\bar 1} {\bar 1} 1 1 1 ), 
&\psi_{26} := (1 1 1 {\bar 1} 1 {\bar 1} {\bar 1} {\bar 1} ), \\
\psi_{27} := (1 1 1 {\bar 1} 1 {\bar 1} 1 1 ), 
&\psi_{28} := (1 1 1 1 {\bar 1} 1 {\bar 1} 1 ), 
&\psi_{29} := (1 1 1 1 {\bar 1} 1 1 {\bar 1} ), \\
\psi_{30} := (1 1 1 1 1 {\bar 1} {\bar 1} 1 ), 
&\psi_{31} := (1 1 1 1 1 {\bar 1} 1 {\bar 1} ), 
\end{array}
\label{eq:saturated_part1}
\end{equation} 
and 
\begin{equation}
\begin{array}{lll}
\psi_{32} := (1 {\bar 1} {\bar 1} 1 0 0 0 0 ), 
&\psi_{33} := (1 {\bar 1} {\bar 1} {\bar 1} 0 0 0 0 ), 
&\psi_{34} := (1 {\bar 1} 0 0 0 0 {\bar 1} 1 ), \\
\psi_{35} := (1 {\bar 1} 0 0 0 0 {\bar 1} {\bar 1} ), 
&\psi_{36} := (1 {\bar 1} 0 0 0 0 1 1 ), 
&\psi_{37} := (1 {\bar 1} 0 0 0 0 1 {\bar 1} ), \\
\psi_{38} := (1 {\bar 1} 1 1 0 0 0 0 ), 
&\psi_{39} := (1 {\bar 1} 1 {\bar 1} 0 0 0 0 ), 
&\psi_{40} := (1 1 {\bar 1} 1 0 0 0 0 ), \\
\psi_{41} := (1 1 {\bar 1} {\bar 1} 0 0 0 0 ), 
&\psi_{42} := (1 1 0 0 0 0 {\bar 1} 1 ), 
&\psi_{43} := (1 1 0 0 0 0 {\bar 1} {\bar 1} ), \\
\psi_{44} := (1 1 0 0 0 0 1 1 ), 
&\psi_{45} := (1 1 0 0 0 0 1 {\bar 1} ), 
&\psi_{46} := (1 1 1 1 0 0 0 0 ), \\
\psi_{47} := (1 1 1 {\bar 1} 0 0 0 0 ), 
&\psi_{48} := (0 0 1 {\bar 1} {\bar 1} 1 0 0 ), 
&\psi_{49} := (0 0 1 1 {\bar 1} 1 0 0 ), \\
\psi_{50} := (0 0 0 0 1 {\bar 1} {\bar 1} 1 ), 
&\psi_{51} := (0 0 0 0 1 {\bar 1} {\bar 1} {\bar 1} ), 
&\psi_{52} := (0 0 0 0 1 {\bar 1} 1 1 ), \\
\psi_{53} := (0 0 0 0 1 {\bar 1} 1 {\bar 1} ), 
&\psi_{54} := (0 0 1 1 1 {\bar 1} 0 0 ), 
&\psi_{55} := (0 0 1 {\bar 1} 1 {\bar 1} 0 0 ), \\
\psi_{56} := (0 0 1 {\bar 1} {\bar 1} {\bar 1} 0 0 ), 
&\psi_{57} := (0 0 1 1 {\bar 1} {\bar 1} 0 0 ), 
&\psi_{58} := (0 0 0 0 1 1 {\bar 1} 1 ), \\
\psi_{59} := (0 0 0 0 1 1 {\bar 1} {\bar 1} ), 
&\psi_{60} := (0 0 0 0 1 1 1 1 ), 
&\psi_{61} := (0 0 0 0 1 1 1 {\bar 1} ), \\
\psi_{62} := (0 0 1 1 1 1 0 0 ), 
&\psi_{63} := (0 0 1 {\bar 1} 1 1 0 0 ).
\end{array}
\label{eq:saturated_part2}
\end{equation} 

\begin{thm}
The collection of rays \eqref{eq:saturated_part1}, \eqref{eq:saturated_part2}, 
\begin{equation*} 
\mathcal{M} := \lbrace \mathbb{C} \psi_{0}, \mathbb{C} \psi_1, \dots, \mathbb{C} \psi_{63} \rbrace
\end{equation*}
is a saturated Kochen-Specker configuration in $\mathcal{H} = \mathbb{C}^{8}$. 
\end{thm}

\emph{Proof.} 
Denote $N = 64$, $d = 8$. 
One may use the following strategy to verify that $\mathcal{M}$ is a KS configuration on a computer. 
First, find all non-empty subsets $I \subseteq [N]$ such that 
$\mathbb{C} \psi_{i}$ is not orthogonal to $\mathbb{C} \psi_{j}$, for any $i, j \in I$. 
The function $f_I: \mathcal{M} \to \lbrace 0, 1 \rbrace$, defined as 
$f_{I} (\mathbb{C} \psi_i) = 1$, if $i \in I$, and 
$f_{I} (\mathbb{C} \psi_j) = 0$, if $j \in [N] \backslash I$, 
is a \emph{candidate} for a Kochen-Specker colouring of $\mathcal{M}$. 
One needs to check, that for every $I$, the collection 
$\mathcal{M} \backslash \lbrace \mathbb{C} \psi_{i} \rbrace_{i \in I}$ 
always contains a tuple of $d$ mutually orthogonal rays.

To generate the tuples of mutually orthogonal rays one can proceed as follows. 
Take $i_0 \in [N]$. 
Then find $i_1 \in [N]$ such that $i_1 > i_0$ and $\mathbb{C} \psi_{i_1} \perp \mathbb{C} \psi_{i_0}$. 
After that find $i_2 \in [N]$ such that $i_2 > i_1$ and $\mathbb{C} \psi_{i_2} \perp \mathbb{C} \psi_{i_1}, \mathbb{C} \psi_{i_0}$, and so on. 
Having a tuple $i_0 < i_1 < \dots < i_{s - 1}$ of mutually orthogonal rays, $1 \leqslant s \leqslant d - 1$, one needs to check 
that there always exists a $j \in [N] \backslash \lbrace i_0, i_1, \dots, i_{s - 1} \rbrace$ such that 
$\mathbb{C} \psi_{j} \perp \mathbb{C} \psi_{k}$, for all $k \in \lbrace i_0, i_1, \dots, i_{s - 1} \rbrace$. 
This establishes the saturation property. 

For a realisation of this method in C look at the supplementary file \texttt{saturated.c}. 
The supplementary file \texttt{vectors.txt} contains a sequence of $64 \times 8 = 512$ numbers 
$a_0, a_1, \dots, a_{511}$ separated by a space, such that $\psi_{i}^{(k)} = a_{i*8 + k}$, for $i \in [64]$, $k \in [8]$.
An analytical way to prove that the configuration $\mathcal{M}$ is of the Kochen-Specker type 
is discussed in the Appendix A. 
The details of how $\mathcal{M}$ had actually been found are given in the Appendix B. 
\qed

\vspace{0.1 true cm}
\emph{Remark.} 
The configuration $\mathcal{M}$ does not admit a colouring compatible with the partition $(7, 1) \vdash 8$, 
but it admits other types of colourings.
Without going into details, one can mention that $\mathcal{M}$ admits colourings 
compatible with the partitions $(6, 2) \vdash 8$ and $(4, 4) \vdash 8$, but not with the partition $(5, 3) \vdash 8$. 
An example of a colouring compatible with $(6, 2) \vdash 8$ is 
an indicator function $h = 1_{\mathcal{A}}$ of $\mathcal{A} = \lbrace \mathbb{C} \psi_i \rbrace_{i \in A}$, where 
\begin{equation*}
A := \lbrace 0, 1, 3, 4, 5, 7, 11, 15, 32, 33, 36, 37, 60, 61, 62, 63 \rbrace, 
\end{equation*}
and an example of 
a colouring compatible with $(4, 4) \vdash 8$ is an indicator function $h = 1_{\mathcal{B}}$ of 
$\mathcal{B} = \lbrace \mathbb{C} \psi_i \rbrace_{i \in B}$, where 
\begin{multline*} 
B := \lbrace  0, 1, 2, 3, 4, 5, 6, 7, 8, 9, 10, 11, 12, 13, 14, 15, \\
32, 33, 34, 35, 36, 37, 38, 39, 56, 57, 58, 59, 60, 61, 62, 63 \rbrace. 
\end{multline*}
For every $U \in \mathcal{P}_{\perp}^{(8)} (\mathcal{M})$ holds: $\# U \cap \mathcal{A} = 2$, and $\# U \cap \mathcal{B} = 4$. 
\hfill $\Diamond$

\vspace{0.1 true cm}
\emph{Remark.} 
It is of interest to point out the following property of the configuration $\mathcal{M}$, $\# \mathcal{M} = 64$. 
If we look at all maximal cliques of the orthogonality graph $\Gamma (\mathcal{M})$ 
then it turns out that the cardinality of an intersection $\# U \cap U'$, as $U, U' \in \mathcal{P}_{\perp}^{(8)} (\mathcal{M})$, $U \not = U'$, 
can be $0, 1, \dots, 6$, but is never $7$. 
At the same time, the configuration $\mathcal{M}$ is saturated, so we conclude: 
\begin{equation} 
\forall W \in \mathcal{P}_{\perp}^{(7)} (\mathcal{M}) \, 
\exists ! U \in \mathcal{P}_{\perp}^{(8)} (\mathcal{M}): U \supseteq W.
\label{eq:Steiner}
\end{equation}
This property is similar to the property required in the definition of a \emph{Steiner system} $S (t, k, n)$, 
$t = 7$, $k = 8$, $n = 64$, 
except that in the case of $S (64, 8, 7)$ arbitrary subsets of cardinality 7 are allowed, while in \eqref{eq:Steiner} one restricts 
to subsets of mutually orthogonal elements. 
\hfill $\Diamond$

\section{Critical Kochen-Specker on three qubits}

In \cite{RuugeFVO1} one has constructed an example of a \emph{saturated} Kochen-Specker configuration in 8-dimensional space. 
This configuration contains 120 rays and extends the Kochen-Specker example found by \cite{KernaghanPeres} (40 rays). 
It turns out that this 120 rays can be perceived as the rays represented by the 240 elements of the $E_8$-root system 
(each ray is represented by a pair of roots $v$ and $-v$). 

In the previous section we have described 
a new \emph{saturated} configuration formed by 64 rays in 8 dimensions. 
It turns out that the number of rays producing the Kochen-Specker property can be reduced 
at the expense of sacrificing the saturation property. 
In this section I describe  
a \emph{critical} subconfiguration  $\mathcal{N} \subseteq \mathcal{M}$ consisting of 36 rays. 
We keep the notation \eqref{eq:saturated_part1}, \eqref{eq:saturated_part2}, for the vectors $\psi_i \in \mathbb{C}^{8}$, $i \in [64]$.

\begin{thm} 
The collection of rays 
$\mathcal{N} := \lbrace 
\mathbb{C} \psi_{I}
\rbrace_{i \in I}$, 
where
\begin{multline} 
I := \lbrace 
2, 3, 4, 9, 12, 13, 14, 15, 16, 19, 21, 23, 24, 25, 26, 27, 29, 30, \\
32, 33, 34, 37, 39, 40, 41, 43, 46, 48, 51, 52, 55, 58, 59, 60, 61, 62 
\rbrace
\label{eq:critical}
\end{multline}
is a critical Kochen-Specker configuration in $\mathcal{H} = \mathbb{C}^{8}$, $\# \mathcal{N} = 36$. 
\end{thm}

\emph{Proof.} 
The fact that $\mathcal{N}$ does not admit a Kochen-Specker colouring 
is checked on a personal computer in analogy with the configuration $\mathcal{M}$. 
One also needs to check that for every $i \in I$, the collection 
$\mathcal{N} \backslash \lbrace \mathbb{C} \psi_{i} \rbrace$ does admit a Kochen-Specker colouring. 

For an implementation in C take the supplementary file \texttt{critical.c}. 
The sequence of numbers $i_0 < i_1 < \dots < i_{35}$ which form the set $I = \lbrace i_{\alpha} \rbrace_{\alpha = 0}^{35}$ 
is written in a supplementary file \texttt{critseq.txt} using space as a separator. 
\qed

Note that precisely half of the vectors $\psi_{i}$ described by \eqref{eq:critical} representing the rays in $\mathcal{N}$ have non-zero coordinates 
$\psi_{i}^{(k)} \in \lbrace \pm 1 \rbrace$, $k \in [8]$, i.e. $i < 32$, while the other half 
comes from the $32 \leqslant i < 64$ part. 
To explain how the configuration $\mathcal{N}$ had actually been found 
we need first to introduce the tropical subconfigurations in $\mathcal{M}$. 
This is done in the next section and the required details are given in the Appendix B.

\section{Tropical Kochen-Specker  on three qubits}

Let $\mathcal{M}$ be a collection of $n$ rays in a $d$-dimensional Euclidean space $\mathcal{H}$. 
Let $(C_{\mathcal{M}}, \wt{C}_{\mathcal{M}})$ denote the \emph{signature} of $\mathcal{M}$. 
Notice that if $\mathcal{M}$ is a single $d$-tuple of mutually orthogonal rays, 
then $C_{\mathcal{M}} (k)$, $k \in \mathbb{Z}_{> 0}$, is a binomial coefficient $C_{d}^{k}$. 
A signature 
is a convenient concept which allows to formulate a \emph{necessary} condition 
meant to distinguish a pair of configurations with non-isomorphic orthogonality graphs. 
If the signatures are different, then the orthogonality graphs cannot be isomorphic. 
At the same time, intuitively, if the the signatures of two configurations coincide, then there is a 
``high chance'' that their orthogonality graphs are isomorphic, although some additional investigation 
is necessary to establish this isomorphism. 

The signature of the critical Kochen-Specker configuration $\mathcal{N}$ described by \eqref{eq:critical} is as follows: 
\begin{equation*} 
\begin{gathered}
(C_{\mathcal{N}} (1), C_{\mathcal{N}} (2), \dots, C_{\mathcal{N}} (8)) = 
(36, 346, 1224, 2063, 1776, 830, 204, 21), \\
(\wt{C}_{\mathcal{N}} (1), \wt{C}_{\mathcal{N}} (2), \wt{C}_{\mathcal{N}} (3), \wt{C}_{\mathcal{N}} (4)) = 
(36, 284, 536, 212), \\
\end{gathered}
\end{equation*}
and $C_{\mathcal{N}} (k) = 0$, if $k > 8$, and $\wt{C}_{\mathcal{N}} (k) = 0$, if $k > 4$.

The signature of the saturated Kochen-Specker configuration $\mathcal{M}$ described by \eqref{eq:saturated_part1}, \eqref{eq:saturated_part2} is as follows: 
\begin{equation*} 
\begin{gathered}
(C_{\mathcal{M}} (1),
\dots, C_{\mathcal{M}} (8)) = 
(64, 992, 5056, 11504, 13312, 8192, 2560, 320), \\
(\wt{C}_{\mathcal{M}} (1), \wt{C}_{\mathcal{M}} (2), \dots, \wt{C}_{\mathcal{M}} (6)) = 
(64, 1024, 4864, 8512, 5632, 1536), \\
\end{gathered}
\end{equation*}
and $C_{\mathcal{M}} (k) = 0$, if $k > 8$, and $\wt{C}_{\mathcal{M}} (k) = 0$, if $k > 6$.

It turns out that one can can compute all tropical subconfigurations in $\mathcal{M}$ and that all of them 
have the same signatures. 
\begin{thm} 
Let $\mathcal{M} = \lbrace \mathbb{C} \psi_{i} \rbrace_{i = 0}^{63}$ be the saturated Kochen-Specker configuration 
in $\mathcal{H} = \mathbb{C}^{d}$, $d = 8$, 
described in \eqref{eq:saturated_part1}, \eqref{eq:saturated_part2}. The following holds: 
\begin{itemize} 
\item[1)] 
The tropical dimension $\dim (\mathcal{M}) = 6$. 

\item[2)] 
The total number of tropical subconfigurations in $\mathcal{M}$ is 32. 
Their signatures coincide.  
\end{itemize}

\end{thm}

\emph{Proof.} 
The computation of the tropical dimension $\dim_{d} (\mathcal{M})$ can be done on a personal computer. 
To generate \emph{all} tropical subconfigurations 
a program written in C needs about a day on a modest machine.

The total number of maximal cliques in the orthogonality graph $\Gamma (\mathcal{M})$ 
is $n = 320$. 
Let us denote these cliques 
$U_0, U_1, \dots, U_{n - 1} \in \mathcal{P}_{\perp}^{(d)} (\mathcal{M})$, $d = 8$. 
A straightforward computation yields the following fact: 
any collection of maximal cliques $U_{i_0}, U_{i_1}, \dots, U_{i_{q - 1}} \in \mathcal{P}_{\perp}^{(d)} (\mathcal{M})$, 
$i_0 < i_1 < \dots < i_{q - 1}$, $i_{\alpha} \in [n]$, $\alpha \in [q]$,
containing five elements or less,  $q \leqslant 5$, admits an anticlique section. 
On the other hand (and this is the longest part of the computation), 
there exist 6-tuples 
$\lbrace i_0 < i_1 < \dots < i_5 \rbrace$ 
such that $\lbrace U_{i_{\alpha}} \rbrace_{\alpha \in [6]}$ does not admit an anticlique section. 
The number of variants of such tuples is $N = 308992$ and we denote the corresponding variants as 
$\lbrace i_{0}^{(\beta)} < i_{1}^{(\beta)} < \dots < i_{5}^{(\beta)} \rbrace$, $\beta \in [N]$. 
The corresponding computations can be found in the supplementary file \texttt{antisect.c}.

It turns out that for every $\beta \in [N]$ holds: 
$\# \cup_{\alpha \in [6]} U_{i_{\alpha}^{(\beta)}} = 48$. 
The dimension of space in our case is $d = 8$, so, since $48 = 6 \times 8$, we immediately conclude 
that every collection $\lbrace U_{i_{\alpha}^{(\beta)}} \rbrace_{\alpha \in [6]}$ is a collection of \emph{mutually disjoint} 
maximal cliques, $\beta \in [N]$. 
Intuitively, one may perceive this observation as an effect of \emph{``repulsion''} of cliques: 
the corresponding configuration tries to be as large as possible. 

The unions $\cup_{\alpha \in [6]} U_{i_{\alpha}^{(\beta)}}$ are Kochen-Specker configurations, 
since any KS colouring of such a union would induce an 
anticlique section of $\lbrace U_{i_{\alpha}^{(\beta)}}\rbrace_{\alpha \in [6]}$, $\beta \in [N]$. 
So our conclusion is as follows: 
$\dim (\mathcal{M}) = 6$.

If we compute the image of the map $[N] \ni \beta \mapsto \cup_{\alpha \in [6]} U_{i_{\alpha}^{(\beta)}} \in \mathcal{P} (\mathcal{M})$, 
then it turns out that it contains only 32 different variants. 
It is straightforward to check that their signatures coincide. 

Write the 32 tropical configurations mentioned as  
$\mathcal{T}_{m} = \lbrace \mathbb{C} \psi_{i} \rbrace_{i \in J_{m}}$, 
where 
$J_{m} = \lbrace j_{m}^{(0)} < j_{m}^{(1)} < \dots < j_{m}^{(47)} \rbrace$, 
for $m \in [32]$. 
The supplementary file \texttt{tropseq.txt} contains a sequence of $32 \times 48 = 1536$ numbers 
$b_0, b_1, \dots, b_{1535}$, 
separated by a space, such that 
$j_{m}^{(s)} = b_{48 m + s}$, $m \in [32]$, $s \in [48]$. 
The only tropical subconfiguration of $\mathcal{M}$ containing $\mathcal{N}$ is $\mathcal{T}_0$. 
The corresponding set $J_{0}$ is as follows:  
\begin{multline} 
J_{0} = \lbrace 
0, 1, 2, 3, 4, 7, 9, 10, 
12, 13, 14, 15, 16, 19, 20, 21, \\ 
22, 23, 24, 25, 26, 27, 29, 30, 
32, 33, 34, 37, 38, 39, 40, 41, \\
43, 44, 46, 47, 48, 50, 51, 52, 
53, 55, 57, 58, 59, 60, 61, 62
\rbrace. 
\label{eq:tropical}
\end{multline}
We have: 
$\mathcal{N} \subseteq \mathcal{T}_{0} \subseteq \mathcal{M}$ and $\# \mathcal{T}_{0} = 48$. 
The computations are implemented in the supplementary file \texttt{tropical.c}.
\qed

The signature of the configuration $\mathcal{T} = \mathcal{T}_{0}$ described by \eqref{eq:tropical} 
in the proof of the theorem is of the shape: 
\begin{equation*} 
\begin{gathered}
(C_{\mathcal{T}} (1), C_{\mathcal{T}} (2), \dots, C_{\mathcal{T}} (8)) = 
(48, 600, 2752, 6096, 7008, 4304, 1344, 168), \\
(\wt{C}_{\mathcal{T}} (1), \wt{C}_{\mathcal{T}} (2), \dots, \wt{C}_{\mathcal{T}} (5)) = 
(48, 528, 1536, 1312, 384), \\
\end{gathered}
\end{equation*}
and $C_{\mathcal{T}} (k) = 0$, if $k > 8$, and $\wt{C}_{\mathcal{T}} (k) = 0$, if $k > 5$.

\section{Thirty-six rays}
There is another known example of a critical Kochen-Specker type configuration due to \cite{KernaghanPeres}. 
It so happens that it also contains 36 vectors, just like the configuration $\mathcal{N}$ described by \eqref{eq:critical}. 
Are these configurations equivalent or not? 
In other words, are the corresponding orthogonality graphs isomorphic or not?

The critical configuration discovered in \cite{KernaghanPeres} can be described as follows. 
Consider first a configuration 
$\mathcal{T}' := \lbrace \mathbb{C} \varphi_{i} \rbrace_{i \in [40]}$, 
represented by 40 vectors of the shape: 
\begin{equation*} 
\begin{array}{lll}
\varphi_{0} := (1 0 0 0 0 0 0 0 ), 
&\varphi_{1} := (0 1 0 0 0 0 0 0 ), 
&\varphi_{2} := (0 0 1 0 0 0 0 0 ), \\
\varphi_{3} := (0 0 0 1 0 0 0 0 ), 
&\varphi_{4} := (0 0 0 0 1 0 0 0 ), 
&\varphi_{5} := (0 0 0 0 0 1 0 0 ), \\
\varphi_{6} := (0 0 0 0 0 0 1 0 ), 
&\varphi_{7} := (0 0 0 0 0 0 0 1 ), 
&\varphi_{8} := (1 1 1 1 0 0 0 0 ), \\
\varphi_{9} := (1 1 {\bar 1} {\bar 1} 0 0 0 0 ), 
&\varphi_{10} := (1 {\bar 1} 1 {\bar 1} 0 0 0 0 ), 
&\varphi_{11} := (1 {\bar 1} {\bar 1} 1 0 0 0 0 ), \\
\varphi_{12} := (0 0 0 0 1 1 1 1 ), 
&\varphi_{13} := (0 0 0 0 1 1 {\bar 1} {\bar 1} ), 
&\varphi_{14} := (0 0 0 0 1 {\bar 1} 1 {\bar 1} ), \\
\varphi_{15} := (0 0 0 0 1 {\bar 1} {\bar 1} 1 ), 
&\varphi_{16} := (1 1 0 0 1 1 0 0 ), 
&\varphi_{17} := (1 1 0 0 {\bar 1} {\bar 1} 0 0 ), \\
\varphi_{18} := (1 {\bar 1} 0 0 1 {\bar 1} 0 0 ), 
&\varphi_{19} := (1 {\bar 1} 0 0 {\bar 1} 1 0 0 ), 
&\varphi_{20} := (0 0 1 1 0 0 1 1 ), \\
\varphi_{21} := (0 0 1 1 0 0 {\bar 1} {\bar 1} ), 
&\varphi_{22} := (0 0 1 {\bar 1} 0 0 1 {\bar 1} ), 
&\varphi_{23} := (0 0 1 {\bar 1} 0 0 {\bar 1} 1 ), \\
\varphi_{24} := (1 0 1 0 1 0 1 0 ), 
&\varphi_{25} := (1 0 1 0 {\bar 1} 0 {\bar 1} 0 ), 
&\varphi_{26} := (1 0 {\bar 1} 0 1 0 {\bar 1} 0 ), \\
\varphi_{27} := (1 0 {\bar 1} 0 {\bar 1} 0 1 0 ), 
&\varphi_{28} := (0 1 0 1 0 1 0 1 ), 
&\varphi_{29} := (0 1 0 1 0 {\bar 1} 0 {\bar 1} ), \\
\varphi_{30} := (0 1 0 {\bar 1} 0 1 0 {\bar 1} ), 
&\varphi_{31} := (0 1 0 {\bar 1} 0 {\bar 1} 0 1 ), 
&\varphi_{32} := (1 0 0 1 0 1 {\bar 1} 0 ), \\
\varphi_{33} := (1 0 0 {\bar 1} 0 1 1 0 ), 
&\varphi_{34} := (1 0 0 1 0 {\bar 1} 1 0 ), 
&\varphi_{35} := (1 0 0 {\bar 1} 0 {\bar 1} {\bar 1} 0 ), \\
\varphi_{36} := (0 1 1 0 {\bar 1} 0 0 1 ), 
&\varphi_{37} := (0 1 {\bar 1} 0 1 0 0 1 ), 
&\varphi_{38} := (0 {\bar 1} 1 0 1 0 0 1 ), \\
\varphi_{39} := (0 {\bar 1} {\bar 1} 0 {\bar 1} 0 0 1 ). 
\end{array}
\end{equation*}
After that construct another configuration by excluding four vectors: 
\begin{equation*} 
\mathcal{N}' := \lbrace \mathbb{C} \varphi_{i} \rbrace_{i \in [40] \backslash \lbrace 0, 12, 22, 31 \rbrace}. 
\end{equation*}

\begin{prop} 
The configuration $\mathcal{N}'$ is a critical Kochen-Specker configuration on three qubits. 
\end{prop}

\emph{Remark.} 
There is a typo in the original paper \cite{KernaghanPeres}: 
by accident, the authors exclude 
the vector $\varphi_{27} = (1 0 {\bar 1} 0 {\bar 1} 0 1 0 )$ instead of 
$\varphi_{31} = (0 1 0 {\bar 1} 0 {\bar 1} 0 1 )$. 
\hfill $\Diamond$.

\vspace{0.1 true cm} 
The configuration $\mathcal{N}'$ is an analogue of our configuration $\mathcal{N}$, and 
$\mathcal{T}'$ is an analogue of the tropical configuration $\mathcal{T}$. 
A \emph{saturated} extension $\mathcal{M}'$ of $\mathcal{T}'$ has been constructed in \cite{RuugeFVO1}. 
It turns out that $\# \mathcal{M}' = 120$ and that the rays of $\mathcal{M}'$ can be represented by the 
vectors of the $E_8$ root system 
(this is an observation related to the question about the symmetry of $\mathcal{N}'$ stated in \cite{KernaghanPeres}). 
We have: 
\begin{equation*} 
\mathcal{N}' \subset \mathcal{T}' \subset \mathcal{M}'. 
\end{equation*}
The configurations $\mathcal{N}$ and $\mathcal{N}'$ have the same cardinalities, 
\begin{equation*} 
\# \mathcal{N}' = 36, \quad 
\# \mathcal{N} = 36, 
\end{equation*}
but their signatures are different: 
\begin{equation*} 
\begin{gathered}
(C_{\mathcal{N}'} (1), C_{\mathcal{N}'} (2), \dots, C_{\mathcal{N}'} (8)) = 
(36, 374, 1384, 1991, 1120, 416, 96, 11), \\
(\wt{C}_{\mathcal{N}'} (1), \wt{C}_{\mathcal{N}'} (2), \wt{C}_{\mathcal{N}'} (3), \wt{C}_{\mathcal{N}'} (4)) = 
(36, 256, 448, 192), \\
\end{gathered}
\end{equation*}
and $C_{\mathcal{N}'} (k) = 0$, if $k > 8$, and $\wt{C}_{\mathcal{N}'} (k) = 0$, if $k > 4$.
Therefore the configuration of rays $\mathcal{N}$ can not be isomorphic to $\mathcal{N}'$. 

The signature of the configuration $\mathcal{T}'$ is of the shape: 
\begin{equation*} 
\begin{gathered}
(C_{\mathcal{T}'} (1), C_{\mathcal{T}'} (2), \dots, C_{\mathcal{T}'} (8)) = 
(40, 460, 1880, 2990, 1880, 780, 200, 25), \\
(\wt{C}_{\mathcal{T}'} (1), \wt{C}_{\mathcal{T}'} (2), \wt{C}_{\mathcal{T}'} (3), \wt{C}_{\mathcal{T}'} (4)) = 
(40, 320, 640, 320), \\
\end{gathered}
\end{equation*}
and $C_{\mathcal{T}'} (k) = 0$, if $k > 8$, and $\wt{C}_{\mathcal{T}'} (k) = 0$, if $k > 4$.

Let us also give (for the reference purposes) the signature of the $E_8$ configuration $\mathcal{M}'$: 
\begin{multline*} 
(C_{\mathcal{M}'} (1), C_{\mathcal{M}'} (2), \dots, C_{\mathcal{M}'} (8)) = \\ = 
(120, 3780, 37800, 122850, 113400, 56700, 16200, 2025), 
\end{multline*}
\begin{multline*} 
(\wt{C}_{\mathcal{M}'} (1), \wt{C}_{\mathcal{M}'} (2), \dots, \wt{C}_{\mathcal{M}'} (8)) = \\ = 
(120, 3360, 31360, 120960, 241920, 241920, 103680, 8640), 
\end{multline*}
and $C_{\mathcal{M}'} (k) = 0$ and $\wt{C}_{\mathcal{M}'} (k) = 0$, if $k > 8$.

Note that it is of interest to search for multi-qubit generalisations of these configurations \cite{HarveyChryssanthacopoulos, PlanatSaniga}. 
In \cite{Ruuge1} one can find an infinite family of Kochen-Specker type 
configurations generalising $\mathcal{T}'$ on any number of qubits $n = 4 m - 1$, $m \geqslant 1$.

%\section{xxx} 
%\section{yyy}
%\section{zzz} 

\section{Conclusion and discussion}

The present paper describes explicitly a new \emph{saturated} Kochen-Speker (KS) configuration of 64 rays 
containing a \emph{critical} configuration of 36 rays on three qubits (8-dimensional space). 
It turns out  that this saturated configuration has quite nice properties which can be studied 
in terms of \emph{tropical} subconfigurations introduced in this paper. 

In this section I make some informal remarks about the subject in general. 
First of all, it is quite natural to expect that the definition of a Kochen-Specker (KS) colouring 
(the latter is a certain function on a configuration with values 0 and 1) 
should have a ``multicoloured'' generalisation. 
Intuitively, one may think that these colours can be perceived as elements of some finite group or a similar algebraic structure. 
From the perspective of pure graph theory such a generalisation is certainly possible. 
The problem is that it is not immediately clear 
how to construct a generalisation which would still be of interest in quantum mechanics, but not just in pure mathematics. 
The approach suggested in the present paper is based on the concept of an \emph{entropy} of a \emph{saturated} configuration. 
The important step is the interpretation of a configuration of rays which admits a Kochen-Specker colouring 
as a configuration with entropy equal to zero. 

At first sight, an extension of a given critical Kochen-Specker configuration to a saturated one is not really 
necessary since it only complicates things, i.e. adds new measuring devices to an experimental set-up as if there is not 
enough trouble with decoherence of quantum states. 
Nonetheless a saturated configuration provides a natural environment where a critical configuration ``lives''. 
A saturated Kochen-Specker configuration is a rather special object with many symmetries and in general one would 
expect many isomorphic copies of a given critical configuration inside it. 
There can be different isomorphism classes of critical configurations. 
Intuitively, a saturated KS configuration, or its orthogonality graph, 
is a discrete analogue of the space of pure states 
of a quantum system. 
One can try 
to mimic quantum mechanics on such finite undirected graphs. 

The objects that exist inside a saturated Kochen-Specker configuration $\mathcal{M}$
(e.g. saturated subconfigurations, critical subconfigurations, etc.) can ``move'': 
if we take an automorphism of the orthogonality graph of $\mathcal{M}$, 
then 
a subgraph in it 
does not need to stay fixed. 
It can be transferred to an isomorphic copy of itself. 
A subconfiguration $\mathcal{N}$ in $\mathcal{M}$ has a \emph{signature} 
(the numbers of cliques and anticliques in the orthogonality graph). 
It is natural to perceive 
the number of maximal cliques 
mentioned in the signature as a ``\emph{capacity}'' of the configuration $\mathcal{N}$. 
Intuitively, the higher is the capacity, the higher is the chance that a configuration does not admit a KS colouring. 
It is tempting to term the number of elements in $\mathcal{N}$ as ``\emph{inductivity}'' of the configuration. 
A pair of subconfigurations $\mathcal{N}_1$ and $\mathcal{N}_2$ in $\mathcal{M}$ can ``interact'': 
if the capacity of $\mathcal{N}_1$ is $n_1$, and the capacity of $\mathcal{N}_2$ is $n_2$, then 
the capacity $n$ of $\mathcal{N}_1 \cup \mathcal{N}_2$ is at least $n_1 + n_2$, but it can be $n > n_1 + n_2$. 
The maximal value of $n$ corresponds to a ``resonance effect'', etc. 

The saturated KS configuration of rays $\mathcal{M}$ constructed in the present paper has $64 = 2^{6}$ elements 
and it corresponds to a system of three qubits. It would be of interest to construct a generalisation of this configuration 
for any number of qubits (the number $64$ suggests that it might be possible). 
It would also be of interest to count all kinds of generalised KS colourings of $\mathcal{M}$ corresponding to non-zero entropies. 
Let's leave it for another paper.

\section*{Appendix A}

It is of interest to 
point out a link between the configurations discussed in the present paper 
and the recent work of M.~Waegell and P.~K.~Aravind \cite{WaegellAravind}, where  
they make an important observation about the algebraic nature  
of a certain class of proofs of the Kochen-Specker theorem. 
Consider the matrices: 
\begin{equation*} 
I = \left( 
\begin{matrix}
1 &0 \\
0 &1
\end{matrix}
\right), \quad 
X = \left( 
\begin{matrix}
0 &1 \\
1 &0
\end{matrix}
\right), \quad 
Y = \left( 
\begin{matrix}
0 &1 \\
-1 &0
\end{matrix}
\right), \quad 
Z = \left( 
\begin{matrix}
1 &0 \\
0 &-1
\end{matrix}
\right).  
\end{equation*}
The matrix $I$ is the $2 \times 2$ identity matrix, 
and  $X$, $Y$, and $Z$, are related to the Pauli matrices as follows: 
$\sigma_1 = X$, $\sigma_2 = - i Y$, $\sigma_3 = Z$. 
Write  
\begin{equation*} 
X^{(0)} := I, \quad 
X^{(1)} := X, \quad 
X^{(2)} := Y, \quad 
X^{(3)} := Z, 
\end{equation*}
and put 
\begin{equation} 
A^{(\alpha)} := X^{(\alpha_0)} \otimes X^{(\alpha_1)} \otimes X^{(\alpha_2)}, 
\label{eq:tensor}
\end{equation}
where $\alpha \in [64] = \lbrace 0, 1, \dots, 63 \rbrace$, 
$\alpha = \alpha_0 + 4 \alpha_1 + 16 \alpha_2$, 
for $\alpha_0, \alpha_1, \alpha_2 \in [4] = \lbrace 0, 1, 2, 3 \rbrace$. 
A rather special property of the collection of matrices $\lbrace A^{(\alpha)} \rbrace_{\alpha \in [64]}$
is that there exist quite many variants to choose 
tuples $\mu_{0} < \mu_{1} < \dots < \mu_{s - 1}$, where $\mu_i \in [64]$, $i \in [s]$, $s  = 3, 4, 7$, 
such that 
\begin{equation} 
[A^{(\mu_i)}, A^{(\mu_j)}] = 0, \quad 
\prod_{k \in [s]} A^{(\mu_k)} = \pm 1, 
\label{eq:tuples}
\end{equation}
where $i, j \in [s]$, and $[-, -]$ denotes the commutator, 
the order of the factors in the product does not matter since the matrices commute. 

Suppose we have a hyper-graph with $n$ vertices labelled by 
$A^{(\nu_i)}$, $\nu_i \in [64]$, 
$\nu_i \not = \nu_j$, if $i \not = j$, $i, j \in [n]$, 
and with hyper-edges of cardinalities 3 or 4. 
Assume that the labels of the vertices in every hyper-edge 
correspond to the tuples of the shape \eqref{eq:tuples}. 
A hyper-edge is termed \emph{negative} if the product of the labels of all its vertices is $- 1$, 
and it is termed \emph{positive} if this product is $+ 1$. 
In \cite{WaegellAravind} it is pointed out that 
every time 
we have a hyper-graph like this  
with an \emph{odd} number of negative hyper-edges, 
then if we have  
a property that every vertex is contained in an \emph{even} number of hyper-edges, 
then this yields immediately a proof of the Kochen-Specker theorem. 
A complete characterization of this class of hyper-graphs is an open mathematical problem. 
In the paper mentioned the authors provide explicitly a series of examples 
of the hyper-graphs they have found. 

We observe now that the saturated Kochen-Specker configuration $\mathcal{M}$ (64 rays) obtained the present paper 
admits a proof from the class \cite{WaegellAravind}. 
At the same time this \emph{configuration of rays} does not underlie 
any of the proofs mentioned in \cite{WaegellAravind}.  
We also 
notice that a straightforward search on a computer 
(i.e. without using any symmetries of the Pauli group)
for a saturated Kochen-Specker configuration with $N = 2^6$ rays 
would require going through $C_{135}^{8} = 2214919483920$
variants (the lower index $135$ in the binomial coefficient $C_{135}^{8}$ 
is the total number of all tuples of the type \eqref{eq:tuples} of 
maximal possible length $s = 7$, and the upper index $8$ should be perceived as 
$N/ d$, where $d = 2^3$ is the dimension of 
the  Euclidian space of three qubits). 

It is a common convention to drop the symbol $\otimes$ in the tensor product \eqref{eq:tensor}. 
If we use the notation $I$, $X$, $Y$, $Z$, then we write, for example, just 
$\mathit{IXX}$ in place of $X^{(0)} \otimes X^{(1)} \otimes X^{(1)}$, 
$\mathit{YZX}$ in place of $X^{(2)} \otimes X^{(3)} \otimes X^{(1)}$, etc. 
Consider a list: 
\begin{equation*} 
\begin{matrix}
0: &\mathit{XII} &\mathit{IXI} &\mathit{XXI} &\mathit{IIZ} &\mathit{XIZ} &\mathit{IXZ} &\mathit{XXZ} \\
1: &\mathit{XII} &\mathit{IXX} &\mathit{XXX} &\mathit{IYY} &\mathit{XYY} &\mathit{IZZ} &\mathit{XZZ} \\
2: &\mathit{IXI} &\mathit{ZIX} &\mathit{ZXX} &\mathit{YIY} &\mathit{YXY} &\mathit{XIZ} &\mathit{XXZ} \\
3: &\mathit{XXI} &\mathit{YYX} &\mathit{ZZX} &\mathit{ZYY} &\mathit{YZY} &\mathit{XIZ} &\mathit{IXZ} \\
4: &\mathit{ZXI} &\mathit{YYI} &\mathit{XZI} &\mathit{IIZ} &\mathit{ZXZ} &\mathit{YYZ} &\mathit{XZZ} \\
5: &\mathit{ZXI} &\mathit{ZIX} &\mathit{IXX} &\mathit{XYY} &\mathit{YZY} &\mathit{YYZ} &\mathit{XZZ} \\
6: &\mathit{YYI} &\mathit{XXX} &\mathit{ZZX} &\mathit{YIY} &\mathit{IYY} &\mathit{ZXZ} &\mathit{XZZ} \\
7: &\mathit{XZI} &\mathit{ZXX} &\mathit{YYX} &\mathit{YXY} &\mathit{ZYY} &\mathit{XIZ} &\mathit{IZZ} \\
\end{matrix}
\end{equation*} 
Every line of this list contains a set of 7 mutually commuting operators. 
Each time there is a set of 8 mutually orthogonal one-dimensional joint eigenspaces (rays) associated to it. 
These sets of rays corresponding to different lines of the list are mutually disjoint, and their union 
yields a set of 64 rays. 
This is precisely the \emph{saturated} KS configuration $\mathcal{M}$ introduced in the paper, if we assume that 
the matrices $A^{(\alpha)}$, $\alpha \in [64]$, act on $x = (x_0, x_1, \dots, x_7)$ as follows: 
$A^{(\alpha)} .x = y$, $y = (y_0, y_1, \dots, y_7)$: 
\begin{equation*} 
y_{i_0 + 2 i_1 + 4 i_2} = \sum_{j_0, j_1, j_2 = 0, 1} 
X_{i_0, j_0}^{(\alpha_0)} X_{i_1, j_1}^{(\alpha_1)} X_{i_2, j_2}^{(\alpha_2)} 
\, x_{j_0 + 2 j_1 + 4 j_2}, 
\end{equation*}
for $i_0, i_1, i_2 = 0, 1$, and 
$\alpha_0 + 4 \alpha_1 + 16 \alpha_2 = \alpha$, where $\alpha_0, \alpha_1, \alpha_2 \in [4]$, 
and $X_{k, l}^{(m)}$ denotes the element of the matrix $X^{(m)}$, $m \in [4]$, 
standing in the $(k + 1)$-th row, $k \in [2]$, and the  $(l + 1)$-th column, $l \in [2]$.

Let us point out a proof that the configuration of rays $\mathcal{M}$ is of the Kochen-Specker type. 
The collection of matrices present in the list above contains 26 elements: 
\begin{multline} 
\mathit{XII}, \mathit{IXI}, \mathit{XXI}, \mathit{ZXI}, \mathit{YYI}, \mathit{XZI}, \mathit{ZIX}, \mathit{IXX}, \mathit{XXX}, \\ 
\mathit{ZXX}, 
\mathit{YYX}, 
\mathit{ZZX}, \mathit{YIY}, \mathit{YXY}, \mathit{IYY}, \mathit{XYY}, \mathit{ZYY}, \mathit{YZY}, \\ 
\mathit{IIZ}, \mathit{XIZ}, \mathit{IXZ}, \mathit{XXZ}, 
\mathit{ZXZ}, \mathit{YYZ}, \mathit{IZZ}, \mathit{XZZ}. 
\label{eq:sat_matrices}
\end{multline}
Consider the following subsets of this collection: 
\begin{equation*} 
\begin{matrix}
\phantom{*} 0: &\mathit{IXI} &\mathit{ZIX} &\mathit{YXY} &\mathit{XIZ} \\
\phantom{*} 1: &\mathit{IXI} &\mathit{ZXX} &\mathit{YIY} &\mathit{XIZ} \\
\phantom{*} 2: &\mathit{ZXI} &\mathit{YYI} &\mathit{IIZ} &\mathit{XZZ} \\
\phantom{*} 3: &\mathit{ZXI} &\mathit{XZI} &\mathit{IIZ} &\mathit{YYZ} \\
\phantom{*} 4: &\mathit{YYI} &\mathit{XXX} &\mathit{YIY} &\mathit{XZZ} \\
\phantom{*} 5: &\mathit{XZI} &\mathit{ZXX} &\mathit{YXY} &\mathit{IZZ} \\
\phantom{*} 6: &\mathit{ZIX} &\mathit{IXX} &\mathit{YYZ} &\mathit{XZZ} \\
*7: &\mathit{IXX} &\mathit{XXX} &\mathit{IZZ} &\mathit{XZZ} \\ 
\end{matrix}
\end{equation*} 
There are 15 matrices present in the list: 
\begin{multline*} 
\mathit{IXI}, \mathit{ZXI}, \mathit{YYI}, \mathit{XZI}, \mathit{ZIX}, \mathit{IXX}, \mathit{XXX}, \\ 
\mathit{ZXX}, \mathit{YIY}, \mathit{YXY}, \mathit{IIZ}, \mathit{XIZ}, \mathit{YYZ}, \mathit{IZZ}, \mathit{XZZ}.  
\end{multline*}
Construct a hyper-graph on 15 vertices labelled with these matrices. 
The 4-tuples in the lines of the list above define 8 hyper-edges: the fist seven are \emph{negative}
and the hyper-edge corresponding to the last line marked with a star is \emph{positive}. 
The matrix $\mathit{IZZ}$ occurs 4 times in a hyper-edge, 
and all other matrices occur twice in a hyper-edge. 
The underlying configuration of rays is, therefore, of the Kochen-Specker type.

The hyper-graph mentioned contains just one positive hyper-edge and all other hyper-edges are negative. 
All hyper-edges are of cardinality 4. 
This is not the only hyper-graph of this shape which corresponds to a proof that $\mathcal{M}$ of the Kochen-Specker type, 
but, perhaps, the most simple one. 
The other hyper-graphs with a single positive hyper-edge 
can be generated on a personal computer as follows. 
There are 24 negative hyper-edges of size 4, and 54 positive hyper-edges of size 4 
which can be constructed from the 26 matrices \eqref{eq:sat_matrices}. 
$2^{24}$ is not a ``huge'' number compared to $2^{54}$. 
Look at all hyper-graphs built only from an \emph{odd} number of negative hyper-edges of size 4. 
Count for each vertex the number of hyper-edges containing it, and  
keep only those hyper-graphs which  
contain exactly 4 vertices 
corresponding to an odd number of hyper-edges. 
Then check if this 4-tuple of vertices forms a \emph{positive} hyper-edge of size 4. 
If this is the case, we obtain a proof that the configuration is of the Kochen-Specker type. 
Since every hyper-graph contains just one \emph{positive} hyper-edge, it is natural to classify these hyper-graphs by 
the tuples of numbers $\lbrace n_k \rbrace_{k \in [24]}$, 
where $n_k$ is the number of vertices contained in exactly $k$ \emph{negative} hyper-edges. 
The computation yields 33 variants of $\lbrace n_k \rbrace_{k \in [24]}$. 
In principle, it would be of interest to describe all isomorphism classes of 
KS proofs for $\mathcal{M}$ corresponding to any number of positive hyper-edges.

\section*{Appendix B} 

In the present paper we have encountered three types of Kochen-Specker configurations: 
saturated (the configuration $\mathcal{M}$, 64 rays), 
tropical (the configuration $\mathcal{T}$, 48 rays), 
critical (the configuration $\mathcal{N}$, 36 rays). 
We have: 
$\mathcal{N} \subset \mathcal{T} \subset \mathcal{M}$. 
The search for a configuration $\mathcal{T}$ has been discussed in the main text. 
Let us say a few words about how to come up with configurations $\mathcal{M}$ and $\mathcal{N}$. 

Look at the saturated extension of the example of M.~Kernaghan and A.~Peres (40 rays) given by the 
$E_{8}$ Kochen-Specker configuration (120 rays). 
These rays can be represented by the following vectors: 
64 vectors of the shape 
\begin{equation*}
(1, \pm 1, \pm 1, \dots, \pm 1), 
\end{equation*}
assuming the number of minus signs is even, 
and 56 vectors of the shape 
\begin{equation*}
(0, \dots, 0, 1, 0, \dots, 0, \pm 1, 0, \dots, 0), 
\end{equation*} 
where $1$ stands in the $i$-th position, and $\pm 1$ stands in the $j$-th position, 
$0 \leqslant i < j \leqslant 7$. 
We select these vectors in such a way that the first non-zero coordinate is 1. 
If, for every vector $v$ in the collection, one takes also $- v$, then the whole set is going to be precisely 
the 240 vectors describing the $E_8$ root system. 
It is natural to ask the following questions: 
\begin{itemize} 
\item[1)] This configuration is a KS configuration of rays in 8-dimensional space (the Euclidean space of three qubits). 
Is it possible to generalise it for $N > 3$ qubits? 

\item[2)] Does the $E_8$ configuration contain a smaller configuration which is saturated and at the same time Kochen-Specker (i.e. is it simple)?
\end{itemize}

Let us look at a subconfiguration $\mathcal{A}$ consisting of 64 rays corresponding to the first 64 vectors mentioned above (i.e. no zeros in the lists of coordinates). 
It turns out that this is a \emph{saturated} subconfiguration, but it is \emph{not} Kochen-Specker. 
The good thing about it is that $64 = 2^{6}$ (a power of two). This number is intuitively better if we speak about $N$ \emph{qubits}. 
This subconfiguration is also special in the following sense: one can represent it as a union of two isomorphic saturated configurations 
with 32 elements each; each of these 32-rays configurations splits into a pair of isomorphic saturated 16-rays configurations; 
each of these 16-rays configurations can be represented as a union of two disjoint 8-tuples of mutually orthogonal rays 
(i.e. as $U_0 \cup U_1$, where $U_0, U_1 \in \mathcal{P}_{\perp}^{(8)} (\mathcal{M})$, $U_0 \cap U_1 = \emptyset$). 

Consider an orthogonal transformation $T: \mathbb{R}^{8} \to \mathbb{R}^{8}$,  
\begin{equation*} 
\begin{gathered}
(1 {\bar 1} {\bar 1} {\bar 1} {\bar 1} {\bar 1} {\bar 1} 1 ) \mapsto (1 {\bar 1} {\bar 1} 1 0 0 0 0 ), \quad 
(1 {\bar 1} {\bar 1} {\bar 1} 1 1 1 {\bar 1} ) \mapsto (1 {\bar 1} 1 {\bar 1} 0 0 0 0 ),\\
(1 {\bar 1} 1 1 {\bar 1} {\bar 1} 1 {\bar 1} ) \mapsto (1 1 {\bar 1} {\bar 1} 0 0 0 0 ), \quad 
(1 {\bar 1} 1 1 1 1 {\bar 1} 1 ) \mapsto (1 1 1 1 0 0 0 0 ),\\
(1 1 {\bar 1} 1 {\bar 1} 1 {\bar 1} {\bar 1} ) \mapsto (0 0 0 0 1 {\bar 1} {\bar 1} 1 ), \quad 
(1 1 {\bar 1} 1 1 {\bar 1} 1 1 ) \mapsto (0 0 0 0 1 {\bar 1} 1 {\bar 1} ),\\
(1 1 1 {\bar 1} {\bar 1} 1 1 1 ) \mapsto (0 0 0 0 1 1 {\bar 1} {\bar 1} ), \quad 
(1 1 1 {\bar 1} 1 {\bar 1} {\bar 1} {\bar 1} ) \mapsto (0 0 0 0 1 1 1 1 ). 
\end{gathered}
\end{equation*}
It turns out that the configuration $\mathcal{B} := \mathcal{A} \cup T (\mathcal{A})$ 
is saturated \emph{and} Kochen-Specker 
(the transformation $T$ is found on a personal computer 
going though the variants of orthonormal bases with vectors having coordinates $0$, $1$, or $- 1$). 
The number of elements in $\mathcal{B}$ is $128 = 2^{7}$ (a power of two). 
It remains to notice that $\mathcal{B}$ splits into two isomorphic copies of a saturated Kochen-Specker configuration of 64 rays. 
One of these copies is precisely the configuration $\mathcal{M}$ described in the present paper. 

The splitting mentioned can be noticed as follows. 
Call the number of elements in $\mathcal{P}_{\perp}^{(d)} (\mathcal{K})$ the \emph{capacity} of a configuration of rays $\mathcal{K}$, 
$d$ is the dimension of space (in our case, $d = 8$). 
Take the configuration $\mathcal{A}$ and look at all pairs $U_0, U_1 \in \mathcal{P}_{\perp}^{(8)} (\mathcal{A})$ 
such that $U_0 \cap U_1 = \emptyset$. It turns out that $\# \mathcal{P}_{\perp}^{(8)} (U_0 \cup U_1)$ can be 2 or 4. 
Select the pairs corresponding to the maximal capacity 4. 
This way we obtain a collection 
$\lbrace W_{i} \rbrace_{i \in [m]}$
of 16-tuples of rays, $\# W_i = 16$, $i \in [m]$, (a computation yields $m = 420$). 
Look now at all disjoint pairs 
$W_{i_0}$, $W_{i_1}$ and try to maximise 
$\# \mathcal{P}_{\perp}^{(8)} (W_{i_0} \cup W_{i_1})$. 
It turns out that, for the configuration $\mathcal{A}$, the capacity of $W_{i_0} \cup W_{i_1}$, in case 
$W_{i_0} \cap W_{i_1} = \emptyset$, can be 8, 16, or 24. 
Select the combinations corresponding to the maximal capacity 24. 
The corresponding computation 
yields a collection of sets $\lbrace Q_{i} \rbrace_{i \in [70]}$, 
$\# Q_i = 32$, and $\# \mathcal{P}_{\perp}^{(8)} (Q_{i}) = 24$, $i \in [70]$. 

We have another isomorphic copy $T (\mathcal{A})$ of the configuration $\mathcal{A}$. 
Construct in a similar way the 70 subsets $\widetilde{Q}_{i} \subset T (\mathcal{A})$, $i \in [70]$, 
such that $\# \widetilde{Q}_i = 32$, and 
$\# \mathcal{P}_{\perp}^{(8)} (\widetilde{Q}_{i}) = 24$, $i \in [70]$. 
The capacity of the configuration $\mathcal{A}$ is 240, 
and it can be represented as $\mathcal{A} = Q_{i_0} \cup Q_{i_1}$, where $i_0, i_1 \in [70]$. 
It turns out that in $\mathcal{A} \cup T (\mathcal{A})$ we can do better: 
there exist $i, j \in [70]$ such that $\# \mathcal{P}_{\perp}^{(8)} (Q_{i} \cup \widetilde{Q}_{j}) = 320$. 
The corresponding computation yields 12 variants of such $(i, j)$. 
One of the configurations $Q_{i} \cup \widetilde{Q}_{j}$ is the configuration $\mathcal{M}$ presented in this paper. 
As we know, it is saturated \emph{and} Kochen-Specker. 
In short, $\mathcal{M}$ is constructed following a \emph{``principle of maximal capacity''}. 

Let us now look at the critical configuration $\mathcal{N}$. 
We start with $\mathcal{T}$ (a tropical subconfiguration of $\mathcal{M}$) 
and obtain $\mathcal{N}$ by a process of deleting rays. 
A similar problem is considered in a recent paper \cite{MegillKresimirWaegellAravindPavicic} where the authors 
try to classify \emph{all} critical subconfigurations of the $H_4$ configuration 
(this is a KS configuration in four dimensions found in \cite{AravindLee-Elkin}). 
It is mentioned that one would need a year on a large computer cluster to compute the instances of them all. 
The problem with this approach is that it does not seem to use the \emph{symmetry} of the problem. 
I do not use the corresponding C libraries, and 
my method exploits actively the concept of a \emph{signature} 
(see the definition in the main text) of a configuration. 
Since the aim is just to generate an example of a small critical subconfiguration 
there is no need for optimisations of the C code: 
a personal computer does the job in a few minutes.

Given a KS configuration of $n$ rays, 
look at all $n$ possibilities of deleting a ray. 
This defines $n$ subconfigurations containing $n - 1$ rays each. 
Keep only those configurations which are not KS colourable. 
Denote their number $m$. 
Compute the signatures of each of the remaining $m$ configurations. 
The observation is that the number of signatures $k$ is typically ``much less'' than $m$. 
Intuitively, a pair of configurations with the same signatures have a high chance to be isomorphic. 
Choose a random representative of each signature 
(for example, the first one encountered). 
This leaves $k$ configurations instead of $m$, and this is much better! 
After that, repeat the procedure described above for each of the chosen $k$ configurations 
selecting again the representatives of signatures. 
Repeat this loop until you are left with only instances of critical subconfigurations.

If we start with the tropical KS configuration $\mathcal{T}$ of 48 rays mentioned, 
then, for example, in the first iteration: $m = 47$ and $k = 2$. 
The number $2$ is ``much less'' than $47$.
It so happens, that after 12 iterations, the configuration $\mathcal{N}$ is the only configuration that ``survives''. 
An implementation in C is in the supplementary file \texttt{generate.c}.  
I have also tested the method on the configuration $\mathcal{T}'$ of M.~Kernaghan and A.~Peres (40 rays in 8 dimensions), 
and it yields the critical configuration $\mathcal{N}'$ of 36 rays pointed out in their paper. 
In short, the observation is that the \emph{``method of signatures''} mentioned 
allows to avoid a rapid growth of the number of variants, 
i.e. the computer does not ``hang up''. 
If we are a little more general (this is not the problem I consider in the present paper) and say 
that we can select representatives of the isomorphism classes of orthogonality graphs of configurations 
instead of representatives of their signatures, then this yields a method to compute all isomorphism classes of 
critical subconfigurations of $\mathcal{M}$.

\end{document}